Review Paper

# Evaluating the anticipated outcomes of MRI seizure image from open-source tool- Prototype approach

Jayanthi Vajiram[1], Aishwarya Senthil[2*], Utkarsh Maurya[3]

[1,] Research Scholar, Vellore Institute of Technology, Chennai, jayanthi.2020@vitstudent.ac.in, utkarsh.maurya2019@vitstudent.ac.in[3]
[2*] Chettinad Hospital and Research Institute, aishwarya.senthil00@gmail.com

**Abstract:** Epileptic Seizure is an abnormal coterminous neuronal exertion in the brain, and it affects nearly 70 million of the world population (Ngugi et al., 2010). The seizures are triggered by abnormal sensitive sensations, movements, loss, or differences in knowledge, and are associated with neurological morbidity. The continuous development in medicine, about 30 percent of cases are treated with antiepileptic drugs (Kwan and Brodie, 2000), only 7 – 8 percent of these cases are undergone surgery (Echauz and Litt,2002) and the epileptogenic brain without a significant functional insufficiency, so the demand of new approaches is apparent. So many open-source neuroimaging tools are used for metabolism checkups and analysis purposes. The scope of open-source tools like MATLAB, Slicer 3D, Brain Suite21a, SPM, and MedCalc are explained in this paper. The MATLAB was used by 60% of the researchers for their image processing and 10% of them use their proprietary software. More than 30% of the researchers use other open-source software tools with their processing techniques for the study of magnetic resonance seizure images. The MATLAB extended the limitation to combine with the process of these Slicer 3D, Brain suite21a, and SPM tools also. The MedCalc tool is used for statistical analysis purposes. This paper mainly focuses on open-source tools and their processing techniques.

## INTRODUCTION

Clinical neuroscience studies are based on neuroimaging and psychiatric. These studies are very important for genetic data processing, cognitive evaluations, and follow-up. Brain imaging describes the colorful ways to either directly or laterally structure and function the neuroimaging system. Neuroradiologists do the interpretation of structural and functional imaging and clinical analysis of the Brain. Structural imaging deals with the nervous system structure and records the opinion of intracranial complaints of excrescence and injury. Functional imaging deals with the metabolic condition's diagnosis and small lesions of any neurological disorder and cognitive psychology exploration with computer interfaces [1]. Imaging is an important part of medical, research, and clinical practice. The various types of techniques have been used to analyze the visual and physiological features in the medical field, a fact which usually impedes the testing and training process [2- 4]. OpenCuda, Theano,Keras,GpuImage,Yolo,Tensorflow,PyTorch,Caffe,EmguCv,Vxl,Gdal,MIScnn,Tracki

ng,Webgazer,Marvin,Korina,FaceForencis,Sketch Transfer, Ron vehicle, ImageNet-A, mobile net, fritz, computer vision annotation tool,3D-Bonet,Reasoning-Rcnn,Steal,Vg-Ae2,Edvr,Corrflow,Funit,mesh R-Cnn,Deepview and BoofCV are the open source tool .The processing tools are Diy filters, standard filters, python tools are Pil, SciKit-Image, and Dataflow Tools are Filterforge.This review is based on some useful open-source software tools applications.

The visualization of medical imaging in MATLAB, Brain suite 2a, Slicer 3D, and SPM. The medical subject has technological limitations of using these open-source tools are presented through some results in this paper. This proves the software tools either process the raw image for further analysis or manage effectively the image data and provide accurate and reliable information. The five phases of Analysis, Development, Design, Implementation, and Evaluation are used as a guideline for building a training performance of these tools [5,6]. The different methods are required to extract features that are then fed to advanced learning algorithms. This paper presents MATLAB-based analytics, compared with other open-source tool applications of Slicer 3D, SPM, and Brainsuite21a.

**MATERIALS AND METHODS**

**Methodology used in Neuroimaging Open-Source Tool**

The automation of multiple datasets from different open-access sites, with different format structures, is framed by the involvement of huge participants, unclear structure, and will need more time to the accurate and usable database. The dataset production comes under data management. The imaging data structure (Gorgolewski et al., 2016) is a standard one that has the storage of multiple behavioral data and neuroimaging modalities with simple file formats and folder structures, and it is accessible by a wide range of researchers. The data are generally used as input for the different open-source tools. These tools will perform the below-given functions to get the output images for clinical analysis.The figure 1 shows the function of the open source output tools.

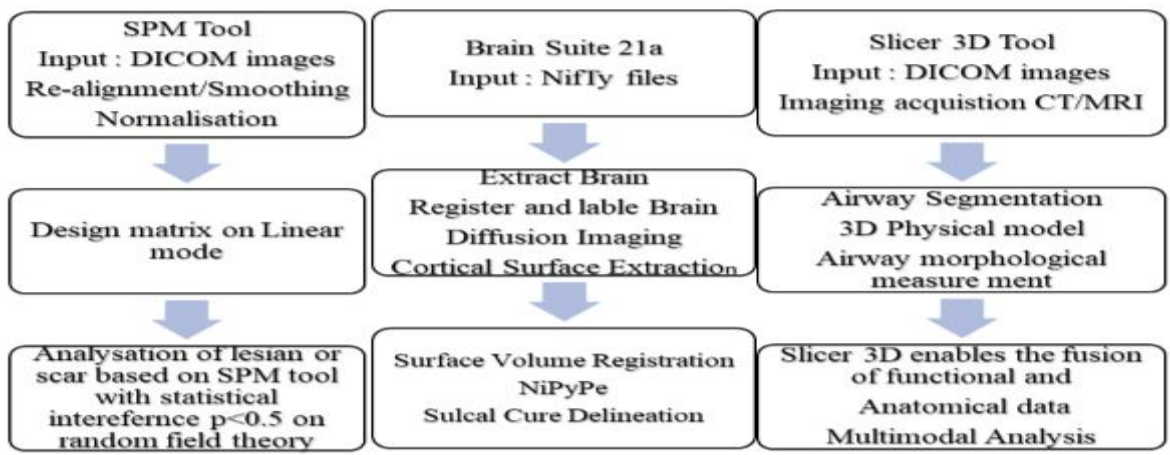

**Figure 1**: Open-source Tools

**MATLAB -Imaging Tool**

MATLAB tool is mainly used in biomedical image processing [7], for filtering, image enhancement, segmentation, and morphological operation [8,9]. The survey of the world wide web service online survey shows that MATLAB stands as one of the best tools among the existing tools through human perception [10,11]. The MATLAB tool covers the qualities of validity, impact, reliability, and practicality. The validity depends on the accurate reflection of the image taken. Reliability concerns the results are accurate, stable, and consistent. Impact concerns the effects and practically concerns the user-friendly tool [12-15]. MATLAB provides extensive data analysis and visualization [16]. The mathematical operation of a two-dimensional photo is image processing [17,18]. The image data is processed with samples and converted into a matrix form and then quantized [19] MATLAB provides easy implementation and algorithmic testing without recompilation [20] as well as easy debugging [21]. It can easily be exploited and enhanced simulation process of both still and video images [22] of the large database based on the algorithms [23]. Figure 2 shows the three steps of the image processing technique [24] of image acquisition, Image analysis, manipulation, and at the end outputting the result. The image of gray matter deficiency of refractory seizure image is used for the analysis of different processing in MATLAB.

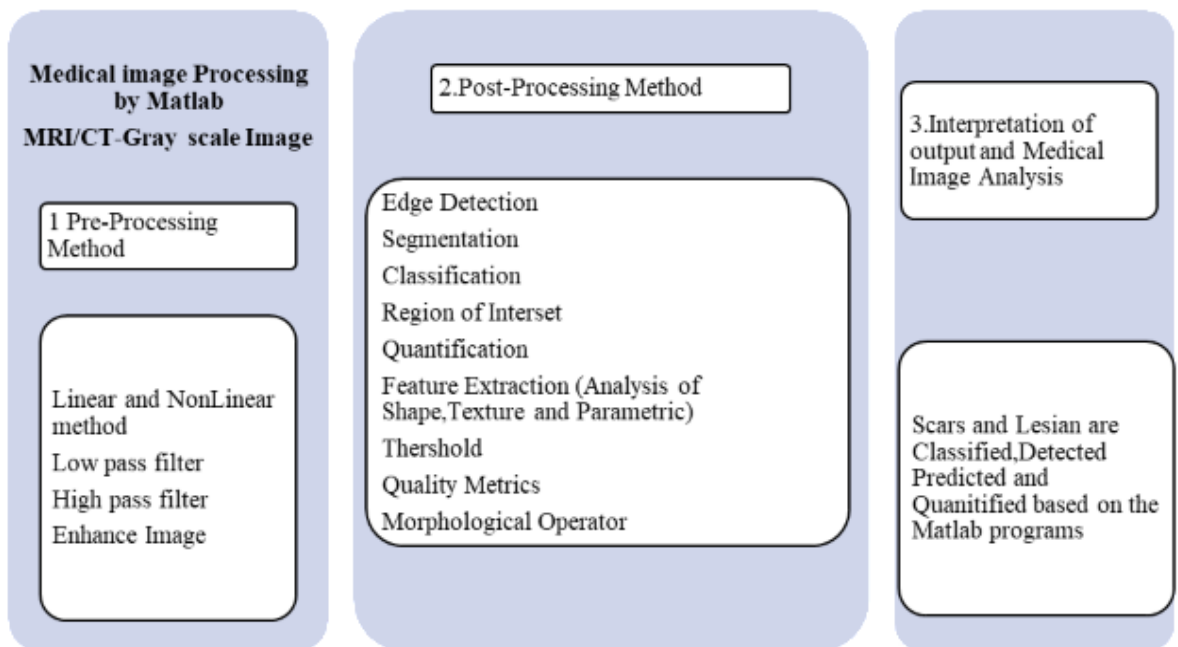

**Figure 2**: Medical Image processing by MATLAB

**The SPM (Statistical Parametric Mapping) -Imaging Tool**

The SPM designed for analysis of brain image sequences of images from different cohorts, or time-series from the same subject. Statistic Parametric Mapping is an open-source software developed for EEG, MRI, and fMRI analysis. It has a Python-MATLAB package of

SPM1d developed by T. Patay for biomechanical research, adapted from neuroimaging. The brain imaging data is processed with the parametric hypothetical image reconstruction method [25]. The volume-based feature uses the general linear model (GLM) measurements of T1 volume pipeline maps with the gray matter (Worsley et al.,2009). The SPM tool was used to compare the seizure-affected images with the healthy control images. SPM tool built to access the large number of volume elements that have spatially correlated voxels (Friston 2007). It processes the sample size calculations and power analysis of (Pataky 2017) pedobarographic videos (Booth et al. 2018) and cortical bone mapping (Li et al. 2009; Poole et al. 2017; Yu et al. 2017) . The hypotheses tests, statistical analysis of a time series of $t$-values allows for better interpretation of data's use Random Field Theory (RFT) (Taylor and Adler 2007)) to mitigate the multiple testing problem, Nyquist criteria, a $p$-value for clusters of statistics ($t$) with the critical threshold ($t^*$) of the Type I error $\alpha = 0.05$. and the cluster has a $p < 0.05$, with null hypothesis $H0$. The available open-source package has both the version of MATLAB and Python [26-27]

**Slicer 3D -Imaging Tool**

Slicer 3D is a multi-platform software package widely used for medical, biomedical, related imaging research. Slicer 3D is used for interactive segmentation and visualization (volume picture) of medical images. Slicer 3D has colorful reslicing options. interpolation, wrapping, masking. Histogram smoothing seems to be elegant colorful enrollment options, powered optimization algorithms. The image template (atlas) data is wrapped in alignment with the input image [28]. It is like the radiology method provides automated segmentation and registration to support the DICOM standards. The slicer extensions are Dcmtk, Qt, Python, and C++. Slicer 3D has DICOM IO support, volume rendering transforms module, and does the process of filtering, registration, segmentation, surface models, diffusion, and image-guided methods. The magnetic resonance images of T1, FLAIR is registered and segmented in a slicer. It shows the contrast T1 images and FLAIR abnormality of seizure identified. The algorithms can be selected based on anatomical site (brain vs registering to an atlas) and check the performance (speed vs accuracy). Slicer 3D retrieves original DICOM images provides automated registration of magnetic resonance images of T1-weighted, T2-weighted, and Flair images [26]. It has SciPy open-source scientific tool, to support slicer SEM, it is incorporated with GIMIAS, Brain, and NiPype [27]. Slicer uses the custom code by ITK to support NRRD formats. The Slicer is from Ontario Consortium for adaptive interventions in radiation oncology (OCAIRO). It is used to detect the mild variation in the abnormal magnetic resonance [28] of seizure images [DICOM images]. The seizure-affected images are compared with the healthy control images.

**The Brain suite 21a -Imaging Tool**

Brain Suite is an open-source software tool used for processing magnetic resonance images (MRI) of the human brain. The MRI analysis sequence in Brain Suite produces individual models of brain structures grounded on T1- ladened MRI of the mortal head. The models produced include brain structures of gross labeling (brainstem, ventricles, mind, cerebellum) charts of towel content in the brain volumes of cerebrospinal fluid (CSF), white matter, gray matter) at each voxel, face models of cranium and crown, and models of cerebral cortex inner and external boundaries to produce a rubric zero face mesh of cerebral cortex in 10-15 twinkles on a 3 GHz computer [32]. Brain suite 21a is used to compare the seizure-affected images with the healthy control images. The Brainstorm was used to import the surfaces of the skull, white/pal, head and read the anatomical label information and Surface volume registration from pial and white matter. It has 15000 samples of head and skull. [33-36].

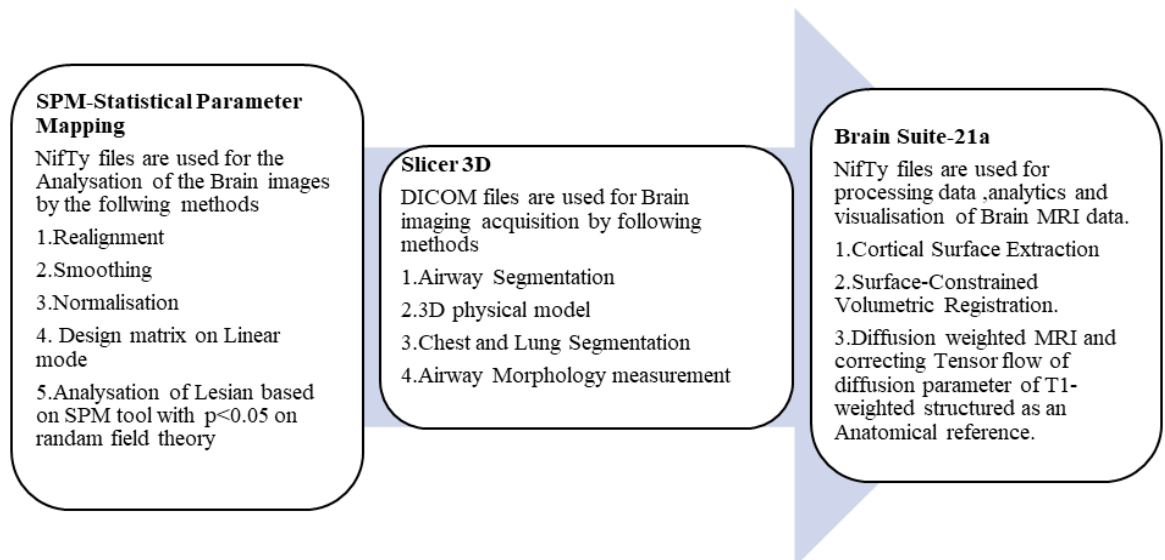

**Figure 3:** The main function of M, Slicer 3D and Brain Suite 21a

**REVIEW OF RELATED WORKS**

The review is based on the related work in MATLAB and open-source software tools. Yang [37] proposed a MATLAB-based pathological information of medical imaging process about a particular ailment. It is used as a visual communication backbone [38]. MATLAB-based image processing techniques are image enhancement, segmentation, and feature extraction [39]. Deepa and Duth [40], Goel et al. [41], proved the RGB color model in picture toolbox for color detection as part of security architecture [42]. Authors in Sahu and Dewangan [43], Mishra et al. [44] used image classification to detect good and bad images. Sharma [45] presented the performance analysis of MATLAB in the biomedical sector for the detection of disease. The images from ultrasound, MRI, and CT scan machines use image processing algorithms to detect diseases and abnormalities. Abdullah et al. [46] used video text extraction for tracking, detection and recognition. Ray et al. [47] used a color-based approach for vehicle detection. The MATLAB edge detection process include Laplacian, Canny, and Sobel for agriculture purposes [48-50]. The edge detection [51], cell growth analysis [52], and grain counting [53] process also incorporated with agriculture. The large number of software packages are available in the market [54]. The methodology of segmentation and registration are behind each tool used to develop the image processing pipelines. The Nipype (Neuroimaging in Python—Pipelines and Interfaces) is an open-source [57] software package (Gorgolewski et al., 2011). The BIDS (Brain imaging data structure) Apps (Gorgolewski et al.,

2017) used for the processing of neuroimaging data, the Nilearn (Abraham et al., 2014) and leverages the sci-kit-learn library (Pedregosa et al., 2011) used for machine learning approaches to neuroimaging data[58]. Statistical analysis of feature extracted images, relies on SPM (Frackowiak et al., 1997) and[59] Surf Stat (Worsley et al., 2009), scikit-learn (Pedregosa et al., 2011) and PyTorch (Paszke et al., 2019). The statistical parametric mapping [60] in Python (Pataky TC. 2012)[61] and biomechanical one-dimensional data with Bayesian inference (Ben Serrien & Jean-Pierre-2019) are analyzed. SPM can read NifTi format, and take image lines like ripples in a pond are called Gibbs Ringing Artifacts. These ripples are large enough caused by moving subject during the scan. SPM is used to get the brain extraction or normalization of the scan images [62].

Slicer 3D supports Windows, Linux, and Mac OS X [63]. Mango software is used to open the patient file sequence as a nii.gz file and drag and apply it into the 3D [64]. Volume's module interface used to adjust window level, color LUT, threshold, and other parameters [65]. The volume rendering module permits the transfer functions of these parameter [66]. The segment editor can adjust the mask, capacity, and slice window outline displays [67]. The eroded skull stripper brain mask serves as dedicated expansion, curvature, and advection terms [68]. The diffusion tensor image projection fibers such as cortical pons, cortical nucleus, spinal thalamus bundle, lateral corpus callosum, cortical cerebellar tract, corticospinal Bundles, spinal cord cerebellum bundle [69]. The diffusion sensor estimation of the DWI converter is used for data scanning [70]. The brain mask is calculated by averaging all baseline images plus segmented tissue voxels (OTSU threshold) [71]. The DWI samples were calculated based on their intensity [72]. The diffusion sensor solar maps were used to measure the different parameter values [73]. Tractography seeding was used to obtain the nerve of the full-beam image [74]. 3D Slicer is a [75] Quantitative Imaging Network (Andriy Fedorov-2012). The 3D anatomical data integration was obtained by 3D optical scanning of CT images [76] for computer-aided implant surgery (Frisardi-2011).

Two-manifold triangular mesh parameterization, used in Brain suite 21a open-source implementation [77]. The head model results are strongly influence the scalp potentials and the inverse source localizations [78]. The model was implemented with eleven tissue types, the structure of the scalp, hard and soft skull bone, CSF, gray and white matter, and other prominent tissues [79], The surface and volume mesh generation techniques used for the right biomedical flow of work [80]. A combination of anisotropic diffusion filtering, edge detection, and mathematical morphology to detect the image of non-brain tissues [81]. Human head simulation is done by Towers [82]. The automated tissue classification of MRI images was done by Van Leemput [83]. The robustness of linear and nonlinear construction of the brain was done by Yerworth [84]. The Brain suite surface is discussed by [85] Shattuck. The automated cortical surface identification tool [86] of Brain suite and its processing methods was analyzed by Leahy.

**System design methodology**

The open-source tool was designed with the following criteria, to input the image file of RGB, intensity, binary and indexed images and used for image type conversions. The concept of color and image color space conversions, standard deviation and mean of images, finding pixel value information, measuring properties of image regions. The image acquisition task is done by the photography of a preloaded image from a memory source [87]. To enhance and restore the quality of RGB images and extract useful information, filtering, edge cropping, gray-level adjustment, and features for graphic display using MATLAB explained by Gerald K. Ijemaru.The MATLAB was used for interpolation and magnification, 2571 types of sharpening, and adjustment of color. In this process, the RGB image is converted to gray-scale, black & white, and binary image by image restoration. The 3D matrix format MxNx3 and matrix of MxN show input matrix represents segments of the different color picture used for RGB image resolution. The Figure 5 shows a block diagram of an image processing technique for the proposed system with different open-source tool is used for analyzing the seizure affected image. The imported image from the image acquisition tool gives altered image as a output. In the SPM voxel-based analyses are used for realigning the data, then the images are subject to non-linear warping for a spatial model for smoothing. The linear model is employed to estimate the temporal model after smoothing process. The SPM test in statistics $t$ or $F$-statistics is based on Random Field Theory and characterizes the responses observed and estimated. SPM is superseded, the ROI and mass-univariate approach. The mass-univariate approach considers images as a multidimensional observation of voxels. find the efficiency of motion pathway of visual cortex, function of the cerebral cortex, brain activation experiments, fMRI time series analysis, unified statistical approach of cerebral activation. The statistical parametric mapping is used for different experimental designs. [88 - 93]

In Slicer 3D the medical images are processed as Digital imaging and communications in medicine (DICOM) are smoothly handled by the 3D Slicer. 3D used to find the volume of the lesions quantitatively from the scan images. It is efficiently used for accessing and interoperating the semantic content in the medical images, medical image markup analysis. This gives clinical decisions confidently based on these data. [94-98]

Brain Suite is designed to process magnetic resonance images (MRI) of the human head. It gives an automatic sequence to extract cortical surface mesh models. This tool is used to register labeled atlas to define anatomical regions of interest and process diffusion imaging data, visualization of data, and can produce interactive maps of regional connectivity. Many experiments are done with this tool like fingerprint analysis of epileptic zone, EEG analysis with two-dimensional support, individual parcellation of resting fMRI, the multi-dimensional approach of magnetic resonance images [99-102]

MDCalc is used to do medical calculations, process algorithms, and scores, assess risk, and create evidence to analyze the condition of patients better and faster.

**SIMULATION RESULTS**

The MATLAB performances based on the image processing technique, employed by algorithms and command codes. The input image of axial T2 weighted grey-matter-heterotopia with refractory seizure shows the results of image processing operations using MATLAB. The seizure affected, DICOM images are converted to Nifty and compared with the healthy control images with the Slicer3D, SPM, and Brainsuite21a tools to analyze the performance of the image.

**MATLAB Simulation Results**

The experimental use of MATLAB algorithms is used to isolate the desired object from the image in order to perform object analysis for clinical use. Figure 4 shows the skull stripping methods explained through the MATLAB code. Figure 5 shows the enhancement method of the image. Figure 6 of input image of axial T2 weighted grey-matter-heterotopia with refractory seizure shows the results of the various image processing operations using the MATLAB. The figure 6a shows the point-detection method, the (x, y) detected points in an image where the mask is centered. It has two masks, in that one mask is associated with a line in the direction of another mask. Each pixel is computed by centering the mask on the pixel location is the output response of the mask. This is used to detect gray isolated spots which is different from its neighbors in an image. The figure 6b and figure 6c shows the Otsu's thresholding method involves iterating of all threshold values to calculate and measure the pixel level spread at each side of the threshold. This method uses the intensity-based histogram result. The figure 6d to figure 6g shows the edge based active contour to do the segregation of the pixels of interest from the image with energy forces and constraints. The algorithms used are Prewitt, Roberts, Sobel, Canny, and fuzzy logic methods. The feature extraction technique uses the Hough transform. It is used to find imperfect instances of objects within a certain class of shapes. The Hough transform takes an input as binary edge points in the edge map, then it is transformed to straight lines are shown figure 6h and figure 6i. The power transform in which the normal distribution of values, which have stable variance and are raised to squared exponential quantity to preserve original images are shown in figure 6j and figure 6k. The figure 6l shows the wavelet transform is a time-frequency analysis, which selects signal characteristics based on frequency band, then it matches the spectrum to improve the time-frequency resolution. It analyzes the abrupt changes in the image [103]. The texture-based segmentation is a part of edge-based relies on edges found by edge detection operators. These edges mark discontinuity in gray levels, color, texture. Whenever the gray level changes, the discontinuity produce edge. Texture segmentation splits an image into different regions and textures with similar pattern of pixel group shown in figure 6m. The figure 6n to figure 6q shows the watershed method for segmentation, change the input image into another image through separating different objects in an image for further analysis

The different feature extraction methods mainly used for character recognition and to define the behavior of an image to improve the clinical proceeding and better understanding are shown in figure 6r to figure 6w. Figure 6r is the feature point detection and description by the BRISK algorithm with scale and rotation invariance to constructs the grayscale relationship in the local image neighborhood, and from the binary feature descriptor. Figure 6s shows the past feature detection method in which data drifts are identified by sequential, model-based, and time distribution methods. Figure 6t shows Harris corner detector is used to extract corners and infer features of an image by computer vision algorithms [104]. Figure 6u shows the Maximally Stable Extremal Regions (MSER) is a feature detector similar to SIFT detector, it extracts an image from co-variant regions. It is stable and the region area variation between different intensity thresholds is not suitable for extreme intensity value changes. Figure 6v shows the detect ORB (Oriented fast and rotated brisk features). Features use the pyramid to produce multiscale features. ORB performs both SIFT and SURF tasks of feature detection. Figure 6w shows to detect SURF (speeded up robust features). It can be used for, image registration, object recognition, classification, or 3D reconstruction [105].

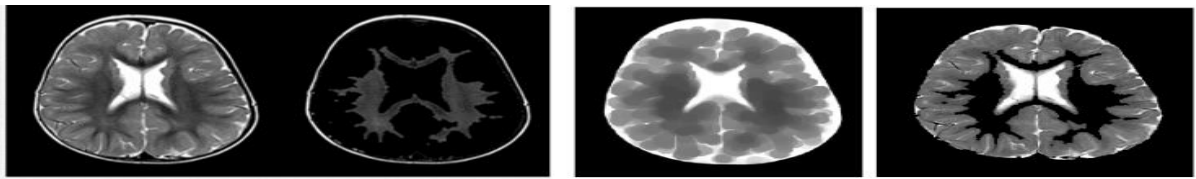

Figure 4: Preprocessing of skull stripping method

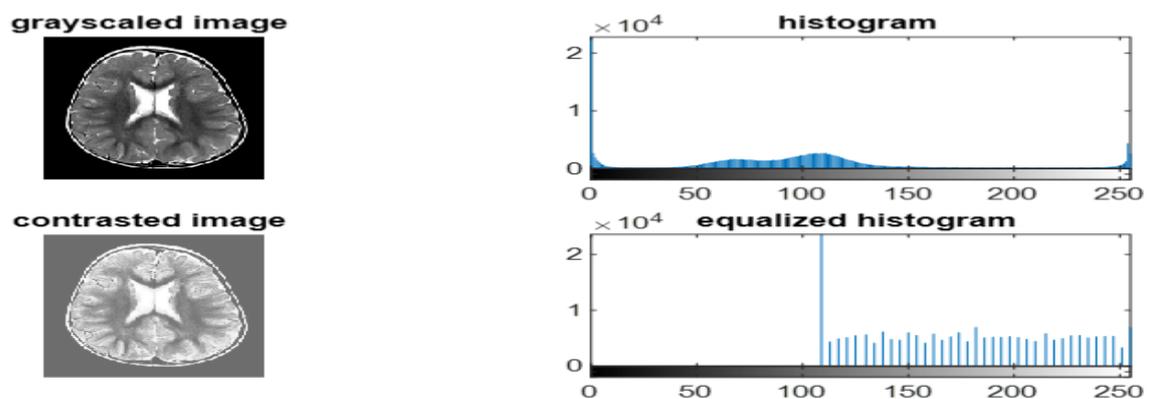

Figure 5: Histogram Enhancement method

**Different Segmentation and Feature Extraction methods**

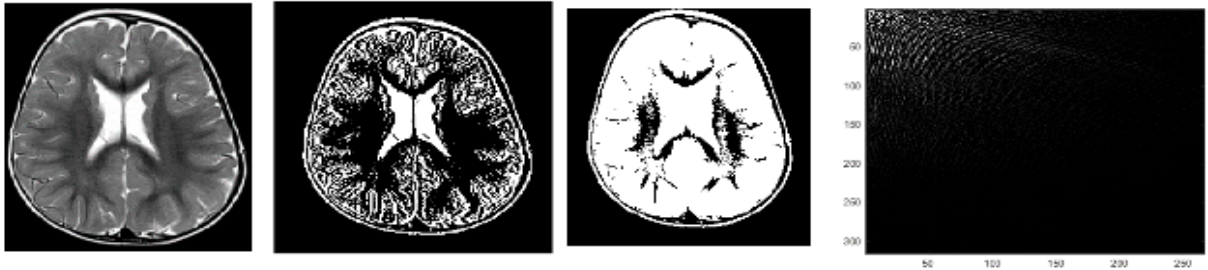

    **Figure :**6          **Figure:** 6a          **Figure :**6b          **Figure :**6c

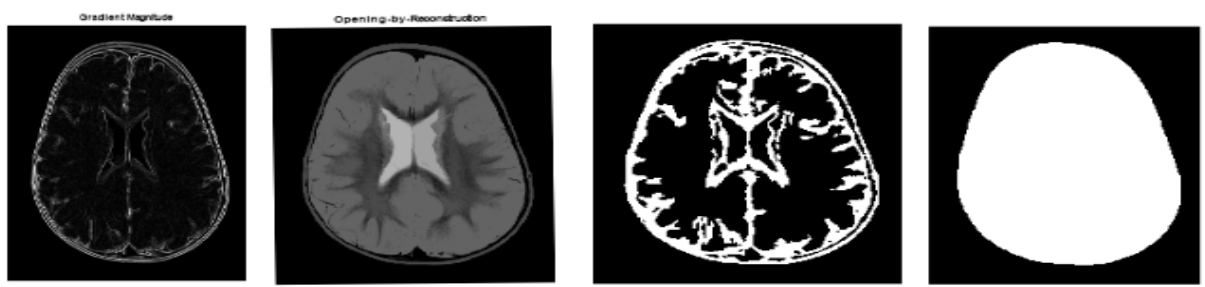

    **Figure :**6d         **Figure :**6e         **Figure :**6f         **Figure:**6g

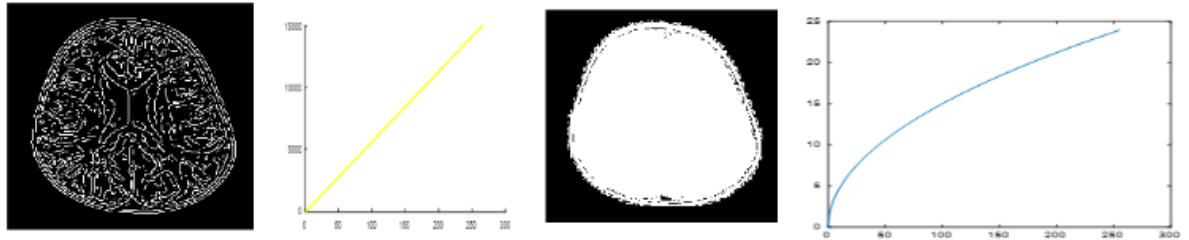

    **Figure :**6h         **Figure :**6i         **Figure :**6j         **Figure :**6k

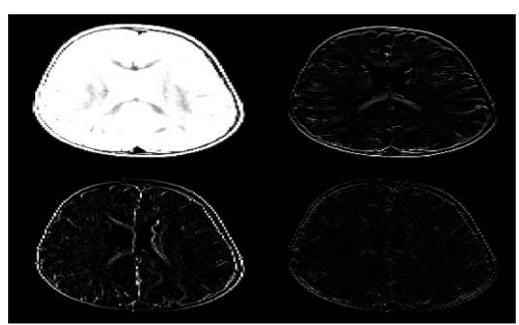 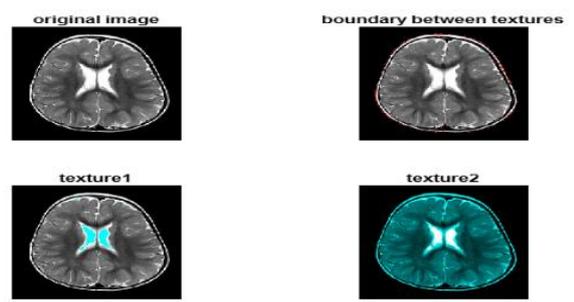

    **Figure :**6l                              **Figure :**6m

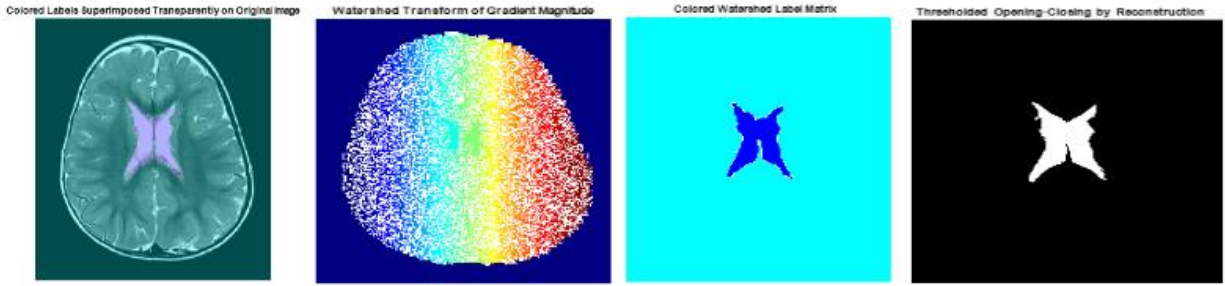

**Figure :**6n      **Figure :**6o      **Figure :**6p      **Figure :**6q

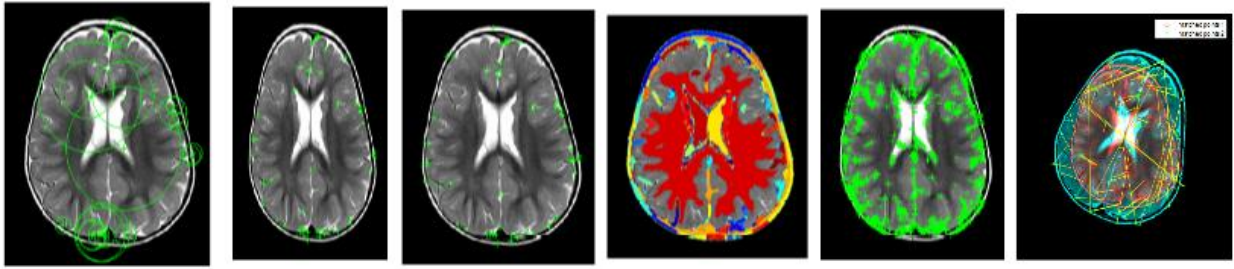

**Figure :**6r    **Figure :**6s    **Figure :**6t    **Figure :**6u    **Figure :**6v    **Figure :**6w

**Figure 6: Input image of Axial T2 weighted grey-matter-heterotopia with Refractory Seizure, Figure 6a: Segmentation using point detection (Singular point detection), Figure 6b: Otsu's Thresholding method, Figure 6c: x and y-axis show the Discrete cosine Transform values of the image, Figure 6d to Figure 6g: Edge-based initial and final level set of active contour model, Figure 6h and Figure 6i: Hough transform and implemented using random to identify the strongest line ,Figure 6j and Figure 6k: power transform approach for segmentation and superimposed regression line, Figure 6l: Wavelet transform, Figure 6m: The Texture based segmentation shows input image, the boundary between textures and texture 1, and texture 2. Figure 6n to Figure6q is watershed segmentation (Figure 6n: Colored labels superimposed transparently on original image, Figure 6m: Watershed transform of Gradient Magnitude, Figure 6o: Colored watershed label matrix and Figure 6p: Threshold opening-closing reconstruction). Figure 6r to Figure 6w is different Feature extraction methods (Figure 6r: To detect BRISK Feature, Figure 6s: To detect Past Feature, Figure 6t: To detect Harris Feature, Figure 6u: To detect MSER Feature and Figure 6v: To detect ORB Feature and Table 1and Table 2 shows the performance metrics of the T2 weighted axial image and Figure 6w explains quality metrics and figure 6x explains the performance metrics of the images.

**Table1**: Performance metrics of T2weighted axial image     **Table2**: Quality metrics

| Performance metrics | T2 weighted Axial Image |
|---|---|
| Accuracy | 0.9221 |
| Jaccard | 0.9745 |
| Structural similarity index | 0.8363 |
| F measure | 0.4431 |
| Specitivity | 0.2177 |
| MCC | 0.2177 |
| Precision | 0.1989 |
| Sensitivity | 0.1443 |
| Dice | 0.1222 |

| Quality metrics | T2 weighted Axial Image |
|---|---|
| Mean Squared Error | 249.7946 |
| Signal to Noise | 32.7938 |
| Peak Signal to Noise | 41.6359 |

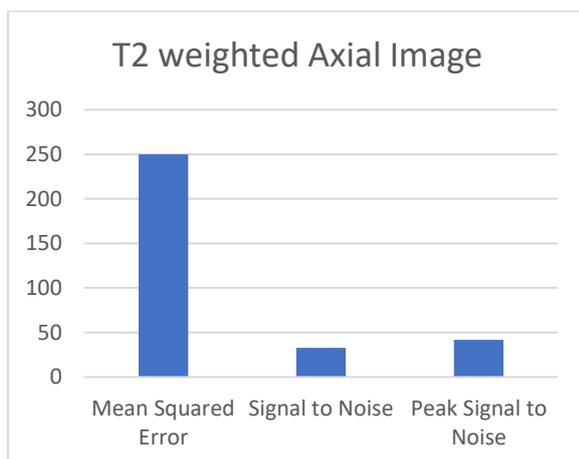

**Figure :**6w

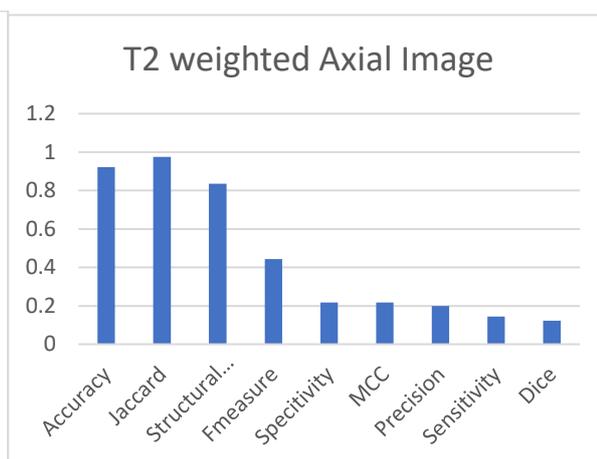

**Figure :**6x

**Results from open-source tool -SPM (Statistical Parametric Mapping)**

Statistical parametric mapping provides a framework that allows the processes of all modalities. The brain's anatomy data to cutting-edge approaches are possible in the SPM.The tool used to model the brain responses, anatomy, rigid body registration, nonlinear registration, segmentation, voxel-based morphometry, random field theory, Bayesian inference, neuronal model, multivariate autoregression and a convolutional model for MRI. The different registration algorithms with gradients, normalization techniques, and entropy correlation coefficient registration are possible in SPM tool [106]. Images collected at higher temporal resolution will convert to a lower spatial resolution, and vice versa. Many of the quality checks based on this tool, which is used for the functional image are the same as with the anatomical image. The normalization and entropy correlation methods are shown in the figure 7a and figure 7b.

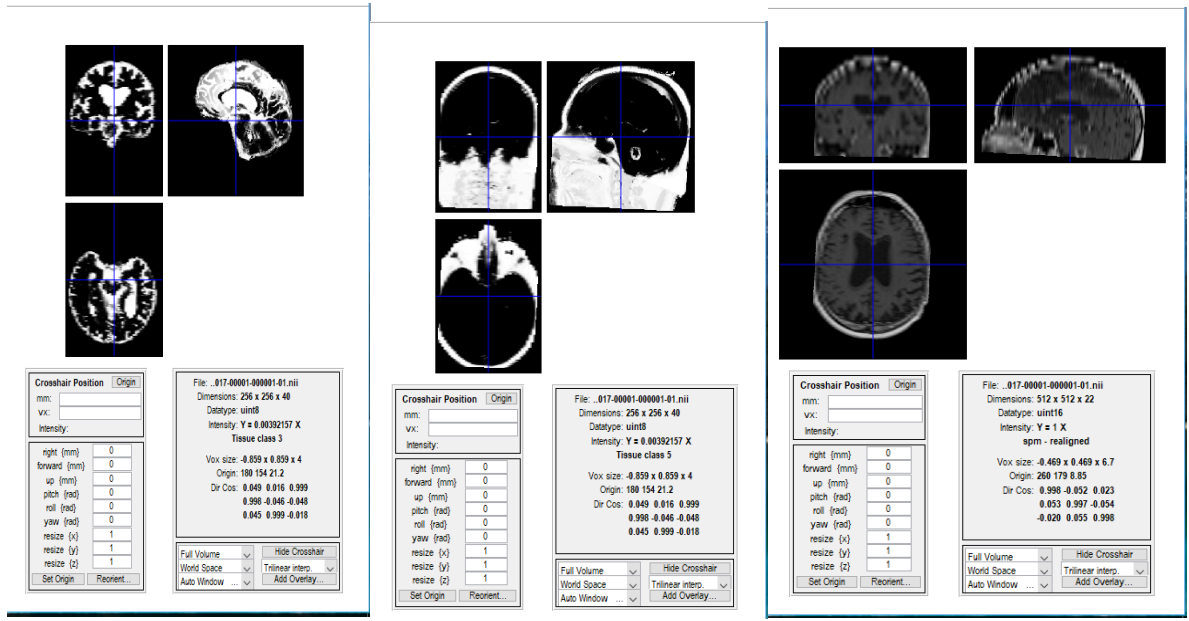

Figure 7a: Seizure MRI,Brain W/WO CONT W/DIFF T1 SAG(Different stages of smooth images) DICOM Image from - SPM)

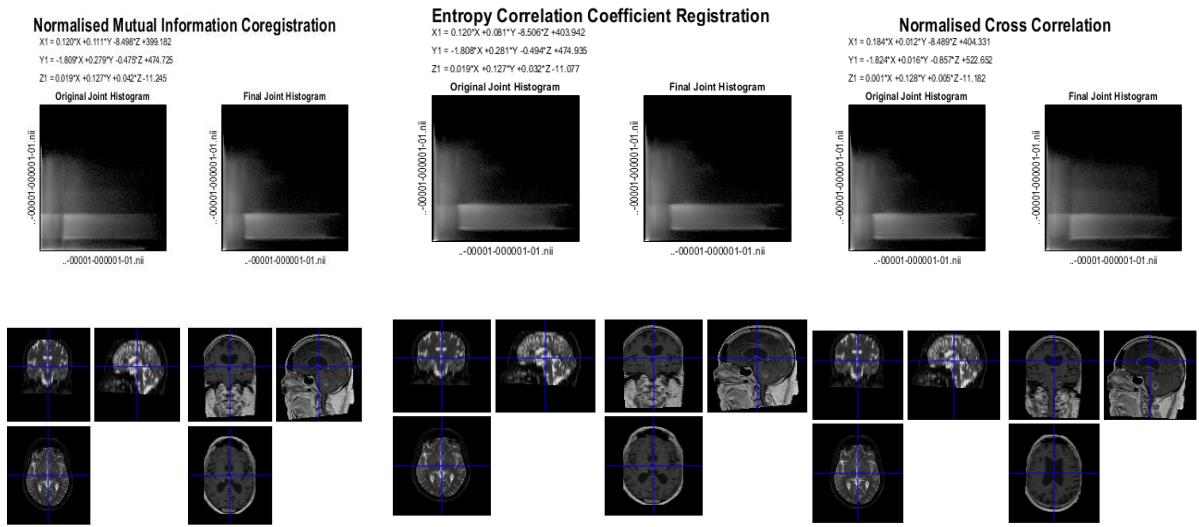

Figure 7b: Normalised ,Coregistration and Entropy of Seizure images compared with healthy control NifTy Image from - SPM)

### Results from open-source tool -Slicer 3D

The slicer markup annotations are in interoperable formats. The different automated tools in slicer 3D assist in delineating different regions of interest (ROI) based on the magnetic resonance images. The figure between 8a and 8d shows the seizure enhancement based on necrosis and edema. The volumetric analysis based on the region of interest to assess the seizure response is either stable or progress.

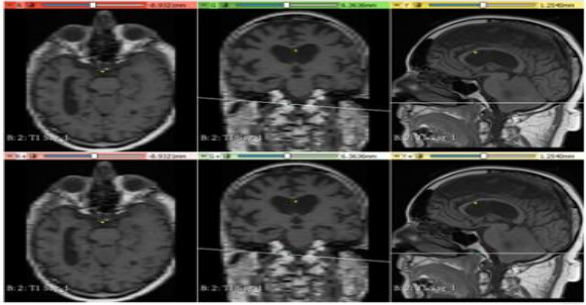

Figure 8a: MRI BRAIN W/Wo CONT/Diff T1-Sagitial Seizure image

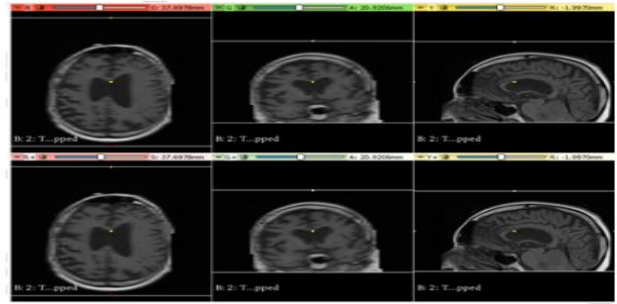

Figure 8b: Crop Volume parameters

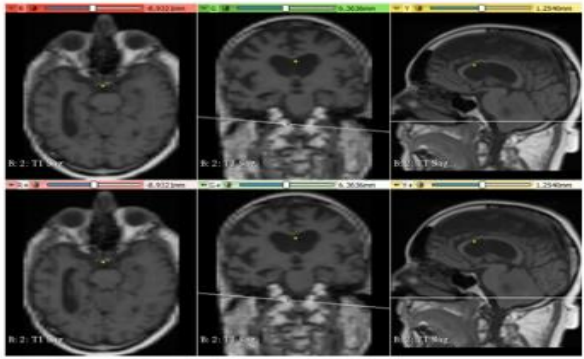

Figure 8c: Gaussian Blur image parameters

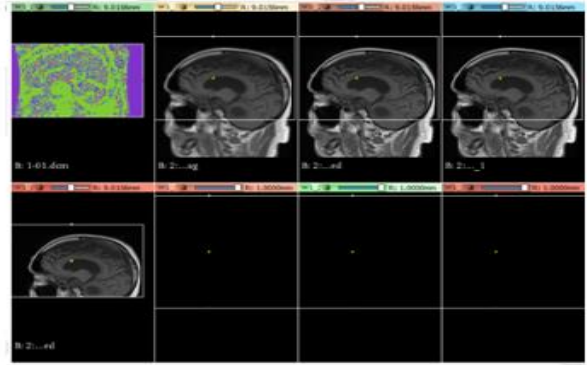

Figure 8d: Volumes are compared

Figure 8a to figure 8d shows the MRI Brain T1 sagittal seizure affected DICOM images and its crop volume, Gaussian blur image details.

### Results from open-source tool – Brainsuite21a

The Brain suite 21a is used for cortical reconstruction. To process the original T1 scan the following process needs to be followed, cortical surface extraction sequence, register and labelling of brain, skull stripping, automated iterations from 5 to 10 or 15. The file T1.svreg.label.nii.gz contains a volume of the subcortical regions. Brainstorm reads these volume labels and tessellates regions, and groups all in a large surface file where the regions are identified in an atlas called structures. The tool will do cortical and volume parcellation. The MRI files were loaded into Brain Suite 21a, and the brain layers were segmented by 'skull stripping (BSD)', which has three iterations value, diffusion constant of 25, and edge constantly of 0.62. The 'skull& scalp' option is used to segment the other layers with thresholds. The figure from 9a to figure 9f shows the result of this tool is based on the seizure-affected image and figure 9g shows the cortical surface extraction of the Hemosiderin affected fMRI images, which are then compared with the healthy control images.

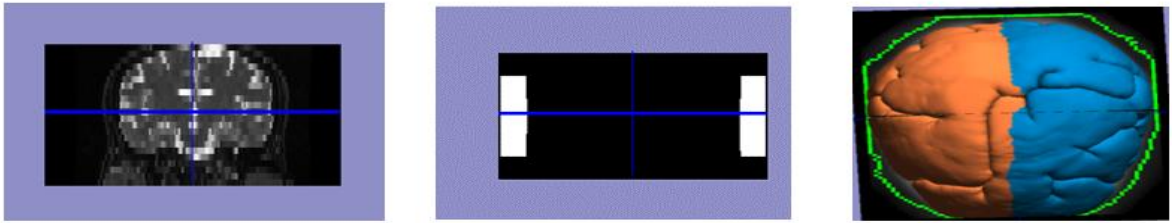

**Figure** 9a: Normal MRI (T1 Sag) (healthy control)   **Figure 9b**: Skull labelling   **Figure** 9c: Cortical surface

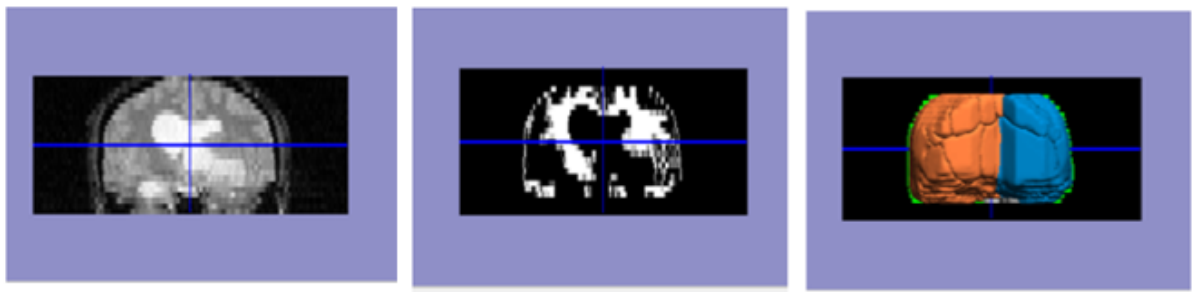

**Figure** 9d: Seizure MRI (T1 Sag )   **Figure** 9e: Gray matter   **Figure** 9f: Cortical surface (Seizure)

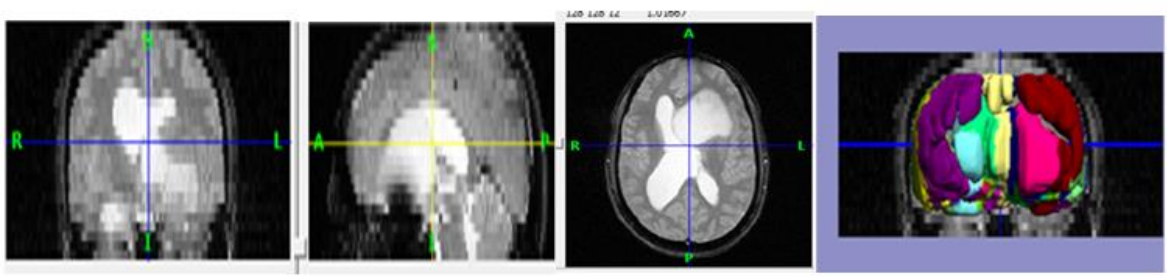

**Figure 9g**: The cortical surface extraction of the fMRI Hemosiderin image

**Results from open-source tool – MedCalc**

**Logistic regression**

It is one of the statistical methods that results the dichotomous or binary variable. It generates the coefficients to predict a logit transformation of the probability of characteristic interest. It chooses parameters that maximize the likelihood of sample values. The figure was drawn based on the two predictor variables of age factor and seizure occurrence and the outcome should be either dependent variable (negative -code 0) or response variable (positive-code 1).

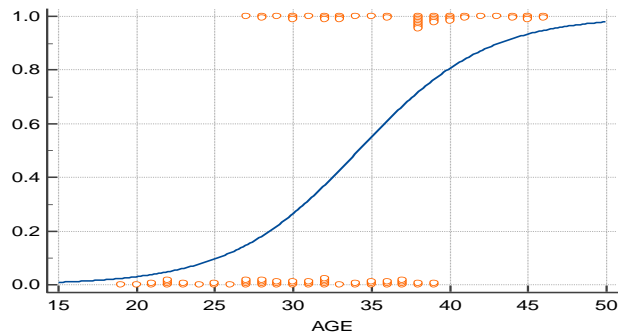

**Figure** 10a: Probability of seizure occurrence based on Age

**ROC Curve analysis**

It represents a sensitivity (true positive)/specificity (false positive) pair corresponding to a particular likelihood ratios and threshold, and positive and negative predictive threshold values. The area under the curve (AUC) with standard error (SE) and 95% confidence interval (CI). ROC curve graph with 95% Confidence Bounds,
- The values, Plot, Partial area under ROC curve, Comparison of precision-recall curves, Interval likelihood ratios, Comparison of partial areas under ROC curves and curve. The Estimation of fixed specificity and sensitivity requires bootstrapping which is a time-consuming technique. This tool gives the difference between the areas under the ROC curves, with 95% confidence interval, standard error and P-value [109]. The graph was captured based on the seizure diagnostic reports. In the following example, there are five predictor variables of seizure diagnosis, gender, and test 1 reports.

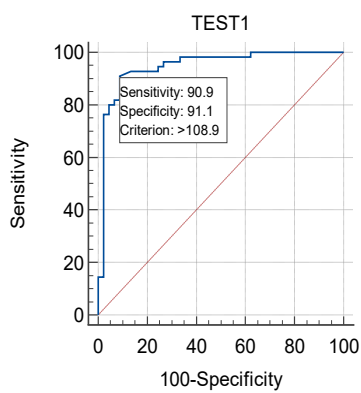 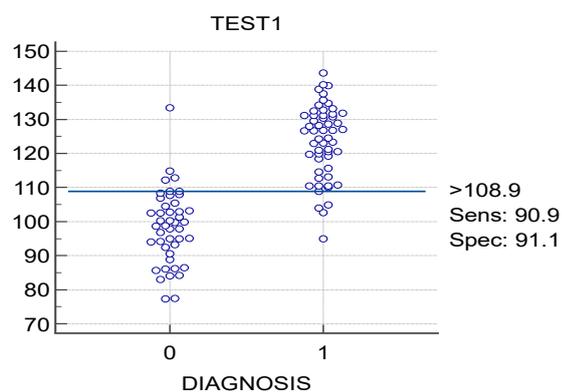

**Figure**10b: ROC curve         **Figure** 10c: Interactive dot diagram

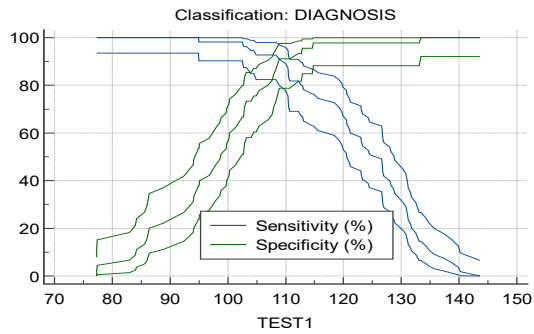

**Figure** 10d: Plot verse criterion value

## CONCLUSIONS

The brain response is captured by simple subtraction, parametric or multifactorial designs, and functional integration, assessed by multivariate analysis framed in terms of effective nonlinear interactions among the input response. An image processing tool package has image utilities and transformation, image filtering, image analysis, image compression, and programming in a data analysis environment. Users can easily rate any software package based on this core parameter, which helps for analyzing and improving the brain-oriented result performance. The algorithm is mainly used to improve the overall performance of image processing. The open-source tool is used for processing different stages of the image like optimization, normalization, registration and volumetric analysis, and realization in medical imaging.

## ACKNOWLEDGEMENT

The authors of this paper would like to offer their deepest gratitude to the School of Electronics Engineering (SENSE), Vellore Institute of Technology - Chennai, for the provision of university resources that proved to be instrumental towards the completion of this undertaking.